\documentclass[12pt,aps,showpacs,eqsecnum,nofootinbib,floatfix]{revtex4}

\usepackage{bbding}
\usepackage{pifont}
\usepackage{subfigure}
\usepackage{epsfig}
\usepackage{array}
\usepackage{amsmath}
\usepackage{amsfonts}
\usepackage{amssymb}
\usepackage{color}
\usepackage{graphicx}
\usepackage{mathrsfs}
\usepackage{multirow}
\usepackage{caption}

\def\be{\begin{equation}}
\def\ee{\end{equation}}
\def\bea{\begin{eqnarray}}
\def\eea{\end{eqnarray}}

\newcommand{\omits}[1]{}

\begin{document}

\title{Microstructure and Continuous Phase Transition of the Gauss-Bonnet AdS Black Hole}
\author{Yun-Zhi Du$^{a,b}$\footnote{Email: duyzh13@lzu.edu.cn}, Hui-Hua Zhao$^{a,b}$, and Li-Chun Zhang$^{a,b}$\footnote{Email: kietemap@126.com(H.-H.Zhao), corresponding author}}

\medskip

\affiliation{\footnotesize$^a$ Department of Physics, Shanxi Datong University, Datong 037009, China\\
\footnotesize$^b$ Institute of Theoretical Physics, Shanxi Datong
University, Datong 037009, China}

\begin{abstract}
The phase transition of the Gauss-Bonnet AdS black hole has the similar property with the van der Waals thermodynamic system. However, it is determined by the Gauss-Bonnet coefficient $\alpha$, not only the horizon radius. Furthermore, the phase transition is not the pure one between a big black hole and a small black hole. With this issue, we introduce a new order parameter to investigate the critical phenomenon and to give the microstructure explanation of the Gauss-Bonnet AdS black hole phase transition. And the critical exponents are also obtained. At the critical point of the Gauss-Bonnet AdS black hole, we reveal the microstructure of the black hole by investigating the thermodynamic geometry. These results perhaps provide some certain help to deeply explore the black hole microscopic structure and to build the quantum gravity.
\end{abstract}

\pacs{04.70.Dy 05.70.Ce}
\maketitle

\section{Introduction}

The investigation of the black hole thermodynamic properties is always the interesting issue of theoretical physics workers. In recent years, people mainly pay attention to the thermodynamic properties of the AdS and dS black holes. Especially, the extended first law equation of black hole thermodynamics was obtained by regarding the cosmological constant in a AdS black hole as the pressure in a ordinary thermodynamic system. Compared the black hole state parameters with the van der Waals (vdW) equation, the critical phenomenons of different black holes were explored by adopting different independent dual parameters. The results showed that the prase transition of a black hole corresponds to the liquid-gas one of a vdW liquid system, and they have the same critical exponents and scalar curvature at the critical phase transition points \cite{Kubiznak2012,Dolan2013,Gunasekaran2012,Sekiwa2006,Cai2013,Hennigar2017a,Wei2015,
Hendi2017,Hendi2017c,Hendi2018,Hendi2017a,Ovguen2018,Oevguen2018,Chaturvedi2017,
Wei2018,Arraut2009,Miao2018,Miao2018a,Mbarek2018,Simovic2019,Ma2018,Ma2017b,Ma2017a,
Zhao2013a,Li2017a,Li2017,Li2017b,Zhao2014,Dayyani2017,Zou2017,Cheng2016,Mir2017,
Banerjee2017,Banerjee2011,Banerjee2012a,Bhattacharya2017,Zeng2017,Dehyadegari2017}.

Although more and more research show that black holes have the common thermodynamic properties with ordinary thermodynamics, the black hole entropy is proportional to the area of horizon radius rather than to the volume. This is a special property of black hole thermodynamic systems. Therefore, the study of the microscopic origin of black hole entropy becomes a a challenge. Among methods of calculating the black hole microscopic state and explaining the microscopic origin of black hole entropy, the string theory offers a natural way. Thereafter, Strominger and Vafa obtained the exact formula of the several supersymmetric black hole entropy by calculating the weakly coupled D-membrane states and extrapolating these results to the black hole phase \cite{Strominger1996}. This method has been applied to other kinds black holes \cite{Emparan2006,Horowitz1996}. Despite of the great achievements, it is valid in the supersymmetric and extreme black holes instead of the Schwarzschild and Kerr black holes. Additional, the black hole microscopic state is still unclear while the black hole entropy can be obtain by different methods.

Since the much consistent for the AdS black hole and the vdW liquid phase transitions, the authors have proposed that the microstructure of black holes is similar to the ordinary thermodynamic system, i.e., black holes are made up of effective black hole molecules at the microscopic scale \cite{Wei2015}. There are some works on the study of the black hole microstructure by introducing the density of black hole molecules and considering the phase transition. Furthermore, the interactions of the black hole molecules are analyzed in Refs. \cite{Miao2018,Miao2018a}. It is clearly that the AdS black hole charge or spin is the necessary condition for the AdS black hole having a similar phase transition with a vdW-like system. That is due to the charge or spin plays a key role in phase transition, which is similar to the effect of magnetization on the phase transition of ferromagnets. Thus in this paper we explore the Gauss-Bonnet AdS black hole microstructure based on this similarity and the Landau continuous phase transition theory. That is not only providing a important window to explore the quantum gravity, but also is of great significance to perfect the thermodynamic geometry theory of black hole.

This work is organized as follows: in Sec. \ref{scheme2}, we present the thermodynamic parameters of the Gauss-Bonnet AdS black hole. For a ordinary thermodynamic system, the phase transition points are the state function of system and are independent with the adoption of the independent dual parameters. In Sec. \ref{scheme3}, we discuss the phase transition of the Gauss-Bonnet AdS black hole for different adoptions of the independent dual parameters by the Maxwell's equal-area law \cite{Li2017,Li2017b}. If one certain adoption of the independent dual parameters will lead to a different phase transition point with other adoptions, the corresponding independent dual parameters are not regarded as black hole independent dual parameters. Therefore, in this part we give the condition of adopting the independent dual parameters to the thermodynamic property of the Gauss-Bonnet AdS black hole. Next \ref{scheme4}, we give the microstructure explanation and phase exponents of the phase transition by introducing a new order parameter $\phi$. In Sec. \ref{scheme5}, the thermodynamic geometry at the critical point is analyzed by the scalar curvature $R$. We also explore the role of the Gauss-Bonnet coefficient $\alpha$ in the phase transition. Finally, a brief summary is given in Sec. \ref{scheme6}.

\section{Gauss-Bonnet Black Hole in AdS Spacetime}
\label{scheme2}
The action of the higher-dimensional Einstein gravity with the Gauss-Bonnet term and cosmological constant $\Lambda=-\frac{6}{l^2}$ in Refs. \cite{Strominger1996,Wei2013,Cai2013} reads
\begin{equation}
I=\frac{1}{16\pi}\int d^dx\sqrt{-g}\left[R-2\Lambda
+\bar\alpha\left(R_{\mu\nu\gamma\delta}R^{\mu\nu\gamma\delta}
-4R_{\mu\nu}R^{\mu\nu}+R^2-4\pi F_{\mu\nu}F^{\mu\nu}\right)\right],
\end{equation}
where the Gauss-Bonnet coefficient $\bar\alpha$ has the dimension with the square length and can be identified with the inverse string tension with positive value. If the theory is incorporated in string theory, thus we shall consider only the case $\bar\alpha>0$. $F_{\mu\nu}$ is the Maxwell field strength defined as $F_{\mu\nu}=-\partial_{\nu} A_{\mu}$ with the vector potential $A_{\mu}$. Note that the Gauss-Bonnet term is a topological one in $d=4$ and has no dynamics in this system. Therefore we will consider the case of $d\geq5$ in the following.

The metric in this system with a static black hole solution is
\begin{eqnarray}
ds^2=-f(r)dt^2+f^{-1}(r)dr^2+r^2h_{ij}dx^idx^j,
\end{eqnarray}
where $h_{ij}dx^idx^j$ represents the line of a $d-2$ dimensional maximal symmetric Einstein space with constant curvature $(d-2)(d-3)k$ and volume $\Sigma_k$. Without loss of the generality, one may take $k=1,~0,~-1$, which are corresponding to the spherical, Ricci fiat, and hyperbolic topology of the black hole horizon respectively. The metric function $f(r)$ was given in Refs. \cite{Wei2013,Cai2013,Xu2014,Wei2014}
\begin{eqnarray}
f(r)=1+\frac{r^2}{2\bar\alpha}\left[1-\sqrt{1+\frac{64\pi\bar\alpha M}{(d-2)\Sigma_kr^{d-1}}-\frac{64\bar\alpha P}{(d-1)(d-2)}}\right].
\end{eqnarray}
Here $M$ represents the ADM mass of the black hole, which is associated with the enthalpy of the system. And $P=-\frac{\Lambda}{8\pi}=\frac{(d-1)(d-2)}{16\pi l^2}$ with the effective AdS curvature radius $l$ is the black hole pressure. In addition, one can use an auxiliary symbol $\alpha=(d-3)(d-4)\bar\alpha$ in order to avoid the verboseness. And we will call the auxiliary symbol $\alpha$ as the Gauss-Bonnet coefficient in the following.

In the present paper we will investigate the phase transition and critical phenomenon for the Gauss-Bonnet AdS black hole in $d=5$ dimensions. The position of the black hole event horizon $r_+$ is determined by a larger root of $f(r_+)=0$. Using the 'Euclidean trick', one have given the black hole temperature, enthalpy, entropy, volume \cite{Miao2018a} as
\begin{eqnarray}
T&=&\frac{8\pi r^3_+P+3r_+}{6\pi(r^2_++2\alpha)},~~~~~~~~~H=M=\frac{3\pi r^2_+}{8}\left(1+\frac{\alpha}{r^2_+}+\frac{4\pi r^2_+P}{3}\right),\nonumber\\
S&=&\frac{\pi^2r^3_+}{2}\left(1+\frac{6\alpha}{r^2_+}\right),~~~~
V=\frac{\pi^2r^4_+}{2}.
\end{eqnarray}
And the equation of state reads
\begin{eqnarray}
P=\frac{3T}{4r_+}\left(1+\frac{2\alpha}{r^2_+}\right)-\frac{3}{8\pi r^2_+}. \label{ES}
\end{eqnarray}
Therefore, the above thermodynamic parameters satisfy the first law \cite{Belhaj2015} as
\begin{eqnarray}
dM=TdS+VdP+\Psi d\alpha
\end{eqnarray}
with the conjugate quantity to the Gauss-Bonnet coefficient $\alpha$
\begin{eqnarray}
\Psi=\left(\frac{\partial M}{\partial\alpha}\right)_{S,P}=\frac{3\pi}{8}-\frac{3\pi^2Tr_+}{4}.
\end{eqnarray}

\section{Equal-Area Law of Gauss-Bonnet AdS Black Hole in Extended Phase Space}
\label{scheme3}
From the equation (\ref{ES}), we know that the equation of state for the Gauss-Bonnet AdS black hole can be transformed to the like-form $f(T,~P,~V,~\alpha)=0$. In the following, we will give the condition of the phase transition with different adoptions of the independent dual parameters $P-V$, $T-S$, $\alpha-\Psi$, and $P-v$.

\subsection{Construction of Equal-Area Law in P-V Phase Diagram}

\begin{figure*}[htb]
\subfigure[$T_0=0.091,~P_0=0.013$]{
\includegraphics[width=5cm,height=4cm]{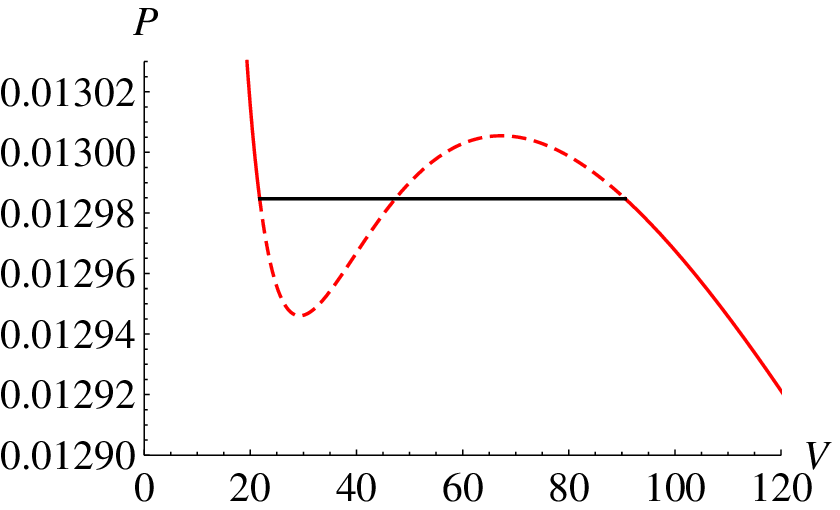}}~
\subfigure[$T_0=0.065,~P_0=0.0065$]{
\includegraphics[width=5cm,height=4cm]{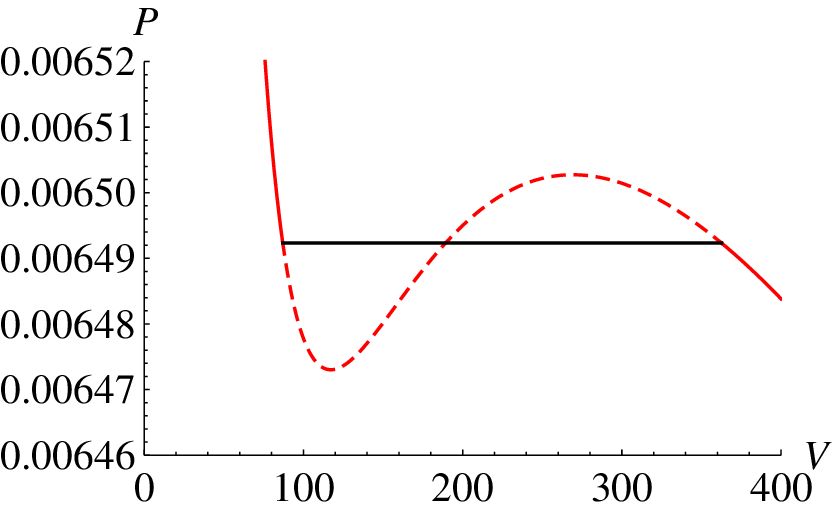}}~
\subfigure[$T_0=0.053,~P_0=0.0043$]{
\includegraphics[width=5cm,height=4cm]{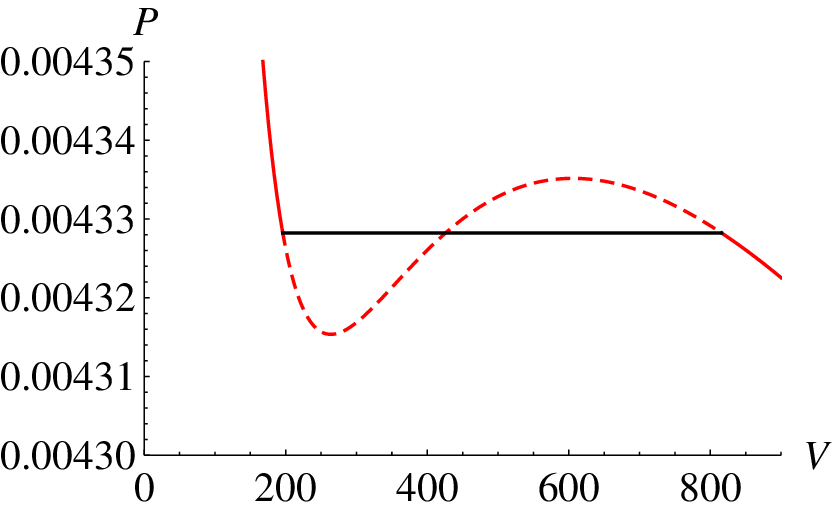}}
\vskip -4mm \caption{The P-V phase diagrams of the Gauss-Bonnet AdS black hole with the parameter $x=\frac{r_1}{r_2}=0.7$, (i.e., $\chi=\frac{T_0}{T_c}=0.994631$). The parameter is set to $\alpha=0.5$ (left), $\alpha=1$ (middle), $\alpha=1.5$ (right)}\label{P-V}
\end{figure*}

For the Gauss-Bonnet AdS black hole with the fixed Gauss-Bonnet coefficient $\alpha$ and the temperature $T_0\leq T_c$ ($T_c$ is the critical temperature), we mark the horizontal and longitudinal coordinates at the boundary of the two-phase coexistence area as $V_1,~V_2$ and $P_0$ in the P-V phase diagram. From the Maxwell's equal-area law \cite{Li2017,Li2017b,Belhaj2015,Miao2018a,Lan2018} ($P_0(V_2-V_1)=\int^{V_2}_{V_1}PdV$), we have the following expresses for this system:
\begin{eqnarray}
P_0=\frac{3T_0}{4r_1}\left(1+\frac{2\alpha}{r_1^2}\right)-\frac{3}{8\pi r_1^2}=\frac{3T_0}{4r_2}\left(1+\frac{2\alpha}{r_2^2}\right)-\frac{3}{8\pi r_2^2},\label{express1}\\
P_0r_2^3(1+x)(1+x^2)=T_0r_2^2(1+x+x^2)+6T_0\alpha-\frac{3}{4\pi}r_2(1+x)
\end{eqnarray}
with $x=r_1/r_2$.
From the above equations, we can obtain
\begin{eqnarray}
r_2^2=\frac{6\alpha}{x},~~~T_0=\frac{3(1+x)}{2\pi r_2(1+4x+x^2)},~~~~
P_0=\frac{3}{4\pi r_2^2(1+4x+x^2)}.\label{express2}
\end{eqnarray}
As $x=1$ (i.e., at the critical point), the critical parameters of this system are
\begin{eqnarray}
r^2_c=6\alpha,~~~V_c=18\pi^2\alpha^2,~~~T_c=\frac{1}{2\pi\sqrt{6\alpha}},~~~~
P_c=\frac{1}{48\pi\alpha},~~~S_c=6\pi^2\alpha\sqrt{6\alpha}.
\end{eqnarray}
For the similarity, by redefining the parameter $\chi\equiv\frac{3(1+x)\sqrt{x}}{1+4x+x^2}$ ($0<\chi\leq1$), the temperature $T_0$ can be rewritten as
\begin{eqnarray}
T_0=\chi T_c=\frac{\chi}{2\pi\sqrt{6\alpha}}.\label{T0}
\end{eqnarray}
For the given temperature $T_0$ and the Gauss-Bonnet coefficient $\alpha$, we can obtain the value of the dimensionless parameter $x$. Then substituting $x$ into the equation (\ref{express2}), the values of $r_2$ (or $r_1$) and the pressure $P_0$ are also known. Based on the classify of the phase transition by Ehrenfest, there is the first-order phase transition for this system with $0<\chi\leq1$. The phase transition curves with the independent dual parameters $P-V$ are shown in Fig. \ref{P-V}.

It is obviously that for the Gauss-Bonnet AdS black hole with a fixed Gauss-Bonnet coefficient $\alpha$ and temperature $T_0$, when the volume $V$ (or the horizon radius $r_+$) is small than $V_1$ (or $r_1$), the phase of Gauss-Bonnet AdS black hole is corresponding to the liquid of a van der Waals system, while it is like the gas of a van der Waals system as $V>V_2$ (or $r_+>r_2$). And the phase is corresponding to the two-phase coexistent of a van der Waals system as $V_1<V<V_2$ (or $r_1<r_+<r_2$).

\subsection{Construction of the Maxwell's Equal-Area Law in T-S phase Diagram}
\begin{figure*}[htb]
\subfigure[$T_0=0.091,~P_0=0.013$]{
\includegraphics[width=5cm,height=4cm]{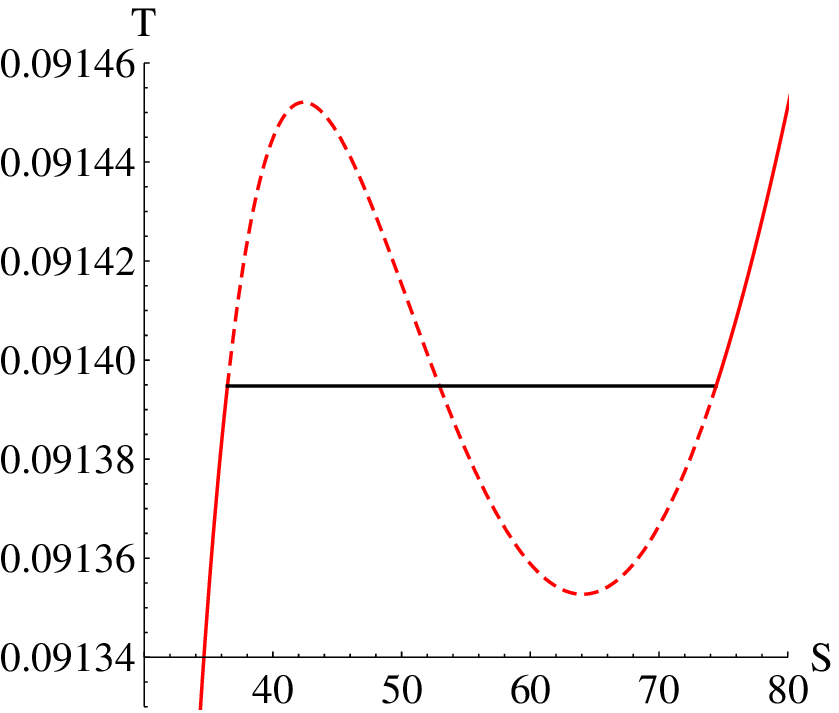}}~
\subfigure[$T_0=0.065,~P_0=0.0065$]{
\includegraphics[width=5cm,height=4cm]{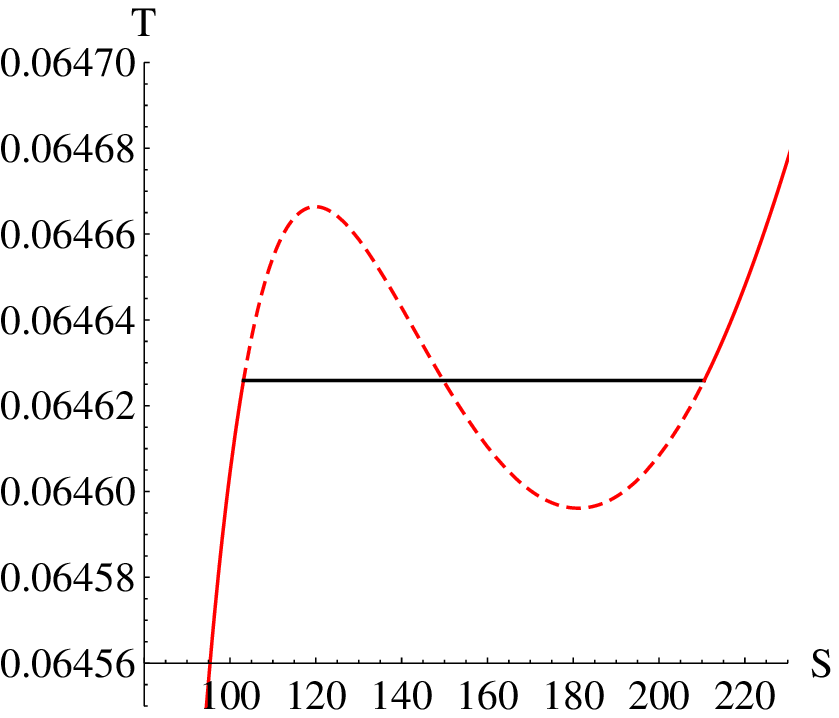}}~
\subfigure[$T_0=0.053,~P_0=0.0043$]{
\includegraphics[width=5cm,height=4cm]{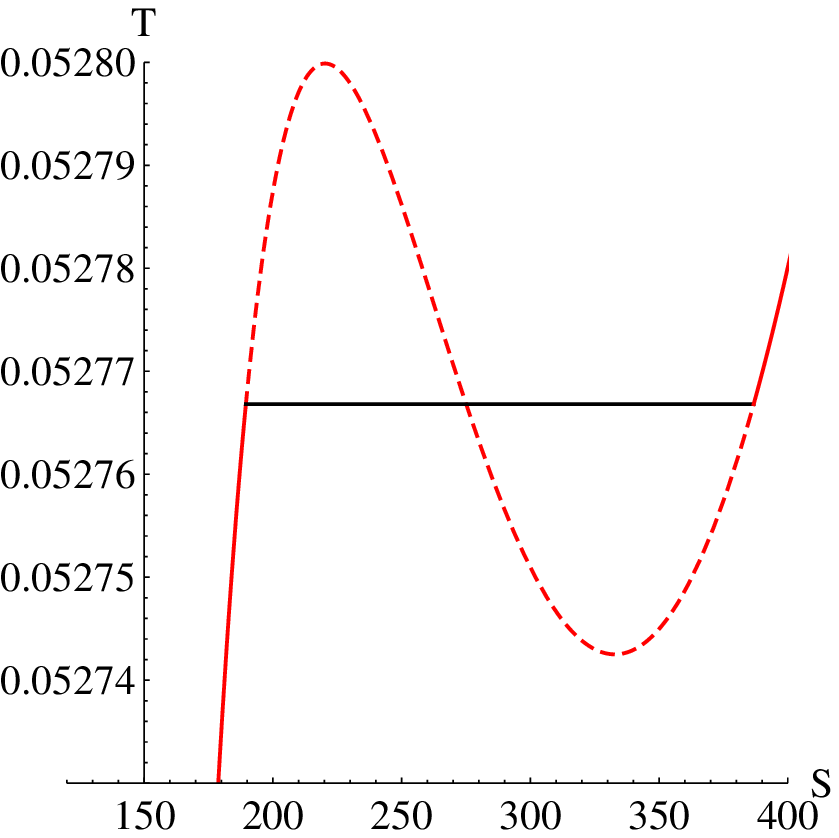}}
\vskip -4mm \caption{The T-S curves of the Gauss-Bonnet AdS black hole phase transition with the parameter $x=\frac{r_1}{r_2}=0.7$. The parameters are set to $\alpha=0.5$ (left), $\alpha=1$ (middle), $\alpha=1.5$ (right)}\label{T-S}
\end{figure*}

For the Gauss-Bonnet black hole thermodynamic system with a certain cosmological constant $l$ in the equilibrium state, we mark the entropies at the boundary of the two-phase coexistence area as $S_1$ and $S_2$, respectively. And the corresponding temperature reads $T_0$, which is less than the critical temperature $T_c$ and is determined by the horizon radius $r_+$. Therefore, from the Maxwell's equal-area law $T_0(S_2-S_1)=\int^{S_2}_{S_1}TdS=\int^{r_2}_{r_1}(8\pi r_+^3P_0+3r_+)dr_+$, we have
\begin{eqnarray}
T_0=\frac{8\pi r_1^3P_0+3r_1}{6\pi(r_1^2+2\alpha)}=\frac{8\pi r_2^3P_0+3r_2}{6\pi(r_2^2+2\alpha)}.
\end{eqnarray}
Note that the solutions of $r_2$, $T_0$, and $P_0$ in the two-phase coexistent state are the same with the equations (\ref{express2}). In the Fig. \ref{T-S}, the T-S phase diagrams are plotted for the different values of pressure $P_0$ and temperature $T_0$ with $\alpha=0.5,~1,~1.5$. It is very clearly that the phase transition point with the same parameter values of $\alpha$ and $x=\frac{r_1}{r_2}$ is consistent with that for the adoption of  the independent dual parameters $P-V$.

It is obviously that for the Gauss-Bonnet AdS black hole with a fixed Gauss-Bonnet coefficient $\alpha$ and pressure, when the entropy $S$ (or the horizon radius $r_+$) is small than $S_1$ (or $r_1$), the phase of Gauss-Bonnet AdS black hole is corresponding to the liquid of a van der Waals system, while it is like the gas of a van der Waals system as $S>S_2$ (or $r_+>r_2$). And the phase is corresponding to the two-phase coexistent of a van der Waals system as $S_1<S<S_2$ (or $r_1<r_+<r_2$).

\subsection{Construction of the Maxwell's Equal-Area Law in $\alpha$-$\Psi$ phase Diagram}

\begin{figure*}[htb]
\subfigure[$T_0=0.091,~P_0=0.013$]{
\includegraphics[width=5cm,height=4cm]{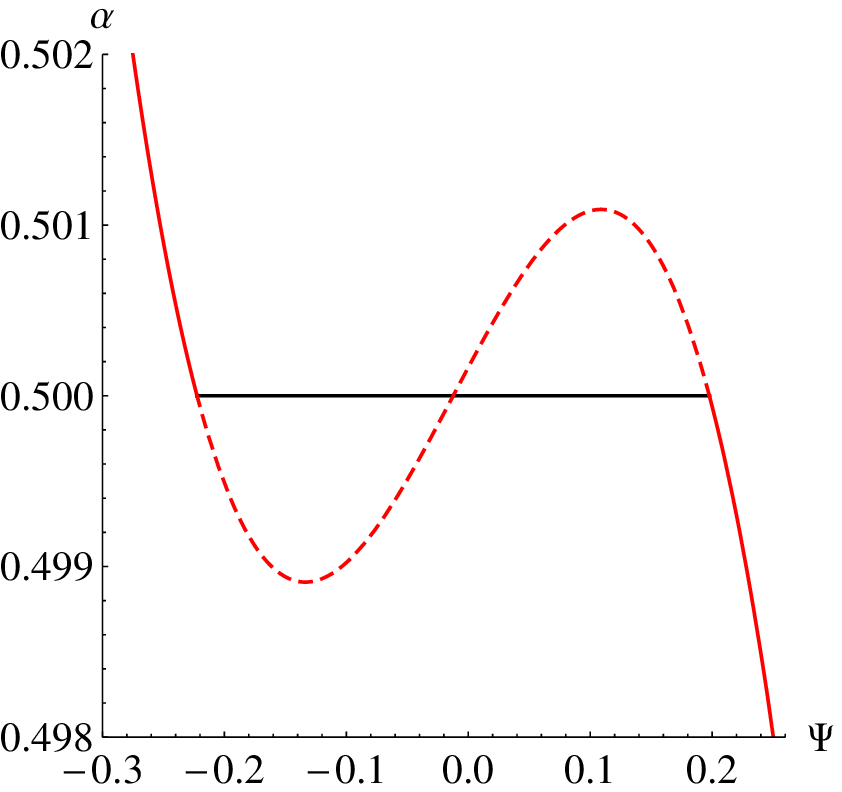}}~
\subfigure[$T_0=0.065,~P_0=0.0065$]{
\includegraphics[width=5cm,height=4cm]{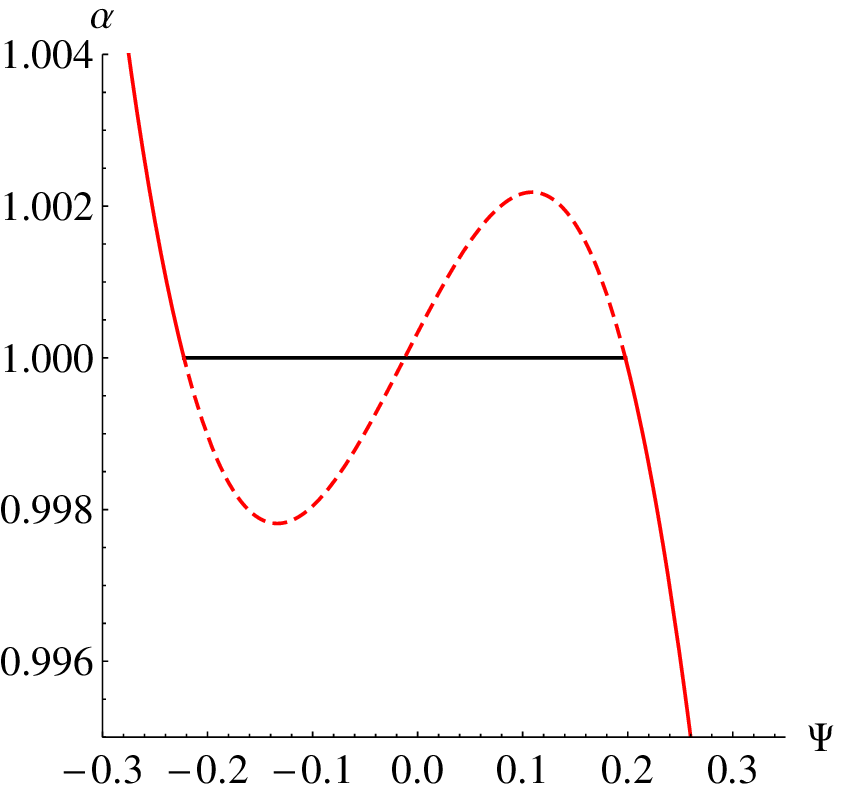}}~
\subfigure[$T_0=0.053,~P_0=0.0043$]{
\includegraphics[width=5cm,height=4cm]{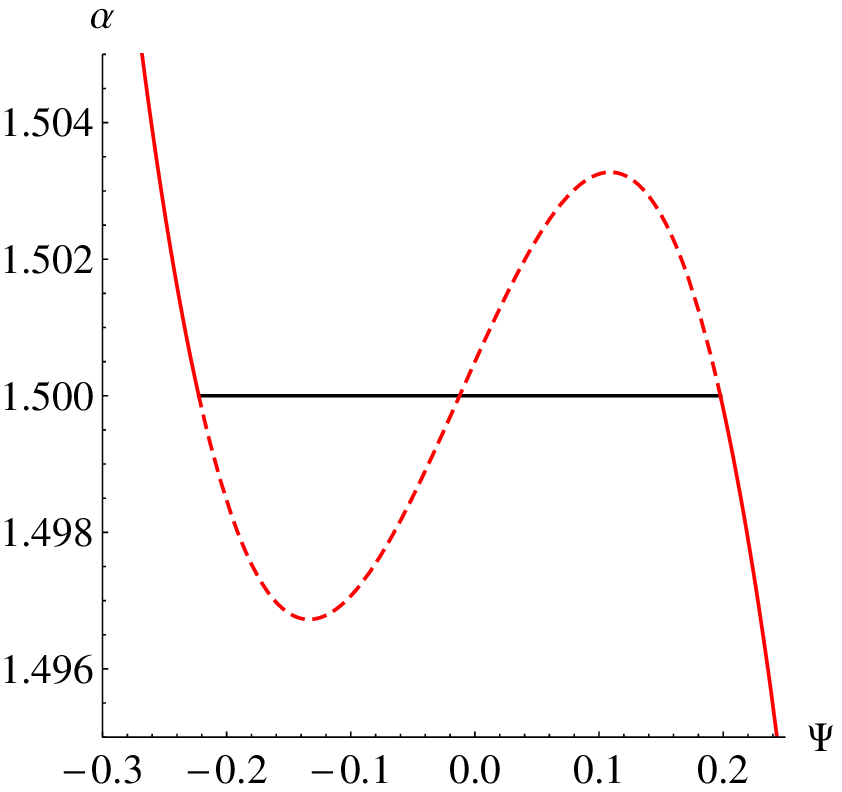}}
\vskip -4mm \caption{The $\alpha$-$\Psi$ curves of the Gauss-Bonnet AdS black hole phase transition with the same temperature, pressure, and the radio parameter $x=\frac{r_1}{r_2}=0.7$. }\label{alpha-Psi}
\end{figure*}

For the Gauss-Bonnet black hole thermodynamic system with a certain cosmological constant $l$ in the equilibrium state, the conjugate quantity $\Psi$ to Gauss-Bonnet coefficient $\alpha$ at the boundary of the two-phase coexistence area are $\Psi_1$ and $\Psi_2$, respectively. And the corresponding Gauss-Bonnet coefficient of the system is $\alpha_0$, which is less than the critical value $\alpha_c$ and is determined by the horizon radius $r_+$. Therefore, from the Maxwell's equal-area law $\alpha_0(\Psi_2-\Psi_1)=\int^{\Psi_2}_{\Psi_1}\alpha d\Psi$, we have
\begin{eqnarray}
\alpha_0=\frac{2P_0r_2^3}{3T_0}+\frac{r_2}{4\pi T_0}-\frac{r_2^2}{2}=\frac{2P_0r_1^3}{3T_0}+\frac{r_1}{4\pi T_0}-\frac{r_1^2}{2}.
\end{eqnarray}
From the above equation, we can obtain
\begin{eqnarray}
r_2^2=\frac{6\alpha_0}{x},~~~T_0=\frac{3(1+x)}{2\pi r_2(1+4x+x^2)},~~~~
P_0=\frac{3}{4\pi r_2^2(1+4x+x^2)}.\label{express3}
\end{eqnarray}
Note that the solutions of $T_0$, and $P_0$ in the two-phase coexistent state are the same with the equation (\ref{express2}). In the Fig. \ref{alpha-Psi}, the $\alpha-\Psi$ phase diagrams are plotted for different values of temperature $T_0$ and pressure $P_0$ with $\alpha_0=0.5,~1,~1.5$. It is very clearly that the phase transition point with the same parameters $\alpha_0$ and $x=\frac{r_1}{r_2}$ is also consistent with that for both the adoptions of $P-V$ and $T-S$.

It is obviously that for the Gauss-Bonnet AdS black hole with the fixed pressure and temperature, when the potential $\Psi$ (or the horizon radius $r_+$) is small than $\Psi_1$ (or $r_1$), the phase of Gauss-Bonnet AdS black hole is corresponding to the liquid of a van der Waals system, while it is like the gas of a van der Waals system as $\Psi>\Psi_2$ (or $r_+>r_2$). And the phase is corresponding to the two-phase coexistent of a van der Waals system as $\Psi_1<\Psi<\Psi_2$ (or $r_1<r_+<r_2$).

\subsection{Construction of Equal-Area Law in P-v phase Diagram}

For the Gauss-Bonnet AdS black hole with the adoption of the dual parameters $P-v$, the volume $v$ at the boundary of the two-phase coexistence area are $v_1$ ($v_1=\frac{4}{3}r_1$) and $v_2$ ($v_2=\frac{4}{3}r_2$), respectively. And the corresponding pressure of the system is $P_0$, which is less than the critical value $P_c$ and is determined by the horizon radius $r_+$. Therefore, from the Maxwell's equal-area law $P_0(v_2-v_1)=\int^{v_2}_{v_1}Pdv$, we have
\begin{eqnarray}
v_2^2&=&-\frac{b}{a},~~~~~~~
T_0=\frac{6x(1+x)\sqrt{-ab}}{32\pi~\alpha(1+x+x^2)a-9\pi b},\\
P_0&=&\frac{3(1+x^2)a}{32\pi\alpha(1+x+x^2)a-9\pi b}-\frac{32\alpha(1+x)(1+x^3)a^2}
{96\pi\alpha(1+x+x^2)ab-27\pi b^2}+\frac{(1+x^2)a}{3\pi x^2b}
\end{eqnarray}
with the relations of $a=486x^5[-2x+(1+x)\ln x+2]$ and $b=144\alpha x^3(1+x)(6+6x^2-12x)$. Note that for the system under the same condition, the form of $r_2$ with this kind adoption of the dual parameters $P-v$ is not the same with that in other adoptions of dual parameters ($P-V,~T-S$, and $\alpha-\Psi$). It implies that the first-order phase transition point of the system with the parameter $v$ will be different from that of the parameter $V$, while the second-order phase transition point is the same. Since in the system with a certain temperature, the location of the first-order phase transition has nothing to do with the adoption of the independent dual parameters, the independent dual parameters $P-v$ are not regarded as the state parameters of the Gauss-Bonnet AdS black hole first-order phase transition.

From the above analyzes for the Gauss-Bonnet black hole phase transition from the Maxwell's equal-area law, we find that:
\begin{itemize}
  \item{From the equation (\ref{T0}), the phase transition is related with the Gauss-Bonnet coefficient $\alpha$ and the horizon radius ratio $x$ ($x=\frac{r_1}{r_2}$), not just only the horizon radius ($r_1$ or $r_2$).}
  \item{For the Gauss-Bonnet AdS black hole with a certain temperature, the independent dual parameters $P-v$ are not regarded as the state parameters of the first-order phase transition.}
\end{itemize}

\section{Microcosmic Explanation of the Gauss-Bonnet AdS Black Hole Phase Transition}
\label{scheme4}

\begin{figure*}[htb]
\begin{center}
\includegraphics[width=7cm,height=6.5cm]{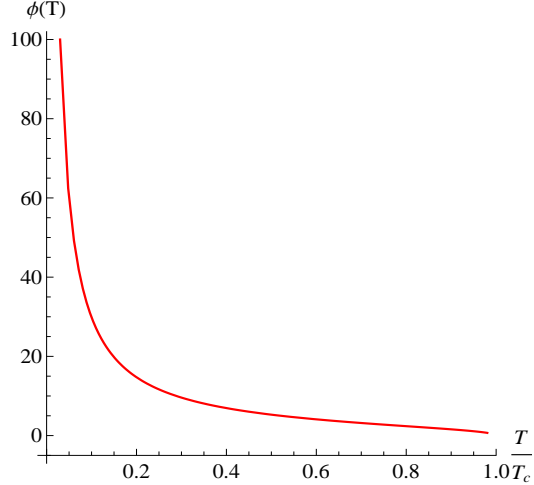}
\end{center}
\vskip -4mm \caption{The $\phi(T)-T/T_c$ curve of the Gauss-Bonnet AdS black hole with $T\leq T_c$.}\label{phi-xi}
\end{figure*}

From the equation (\ref{express2}), we can see that when the Gauss-Bonnet AdS black hole undergoes a phase transition, the values of radio between $\sqrt{\alpha}$ and the horizon radius at the boundary of the two-phase coexistence area have a mutation, i.e.,
\begin{eqnarray}
\phi_1=\frac{\sqrt{\alpha}}{r_1}=\frac{1}{\sqrt{6x}},~~~~
\phi_2=\frac{\sqrt{\alpha}}{r_2}=\sqrt{\frac{x}{6}}.
\end{eqnarray}
Therefore, we introduce the new order parameter $\phi(T)$ as
\begin{eqnarray}
\phi(T)\equiv\frac{\phi_1-\phi_2}{\phi_c}=\frac{1-x}{\sqrt{x}}
=\frac{\Psi_2-\Psi_1}{\chi(\Psi_c-3/8\pi)}\label{phiT}
\end{eqnarray}
with $\phi_c=\frac{1}{\sqrt{6}}$ and $\chi\equiv\frac{3(1+x)\sqrt{x}}{1+4x+x^2}$. Note that $\Psi_c$ is the potential at the critical point, $\Psi_c=\frac{3\pi}{8}-\frac{3\pi^2T_cr_c}{4}$. The plot of the new order parameter $\phi(T)$ with the temperature exponent $\frac{T}{T_c}\leq1$ is given in Fig. \ref{phi-xi}.

The authors in Refs. \cite{Wei2015,Miao2018,Miao2018a} pointed out that the phase transition between a big black hole and a small one is due to the different black hole molecular number densities. Analyzing the effect of the Gauss-Bonnet coefficient $\alpha$ on the phase transition, we reconsider the physical mechanism of the Gauss-Bonnet AdS black hole undergoing a phase transition.

From the Landau continuous phase transition theory, we know that the symmetry of matter will change with the order of matter. Since a black hole has the similar property with a ordinary thermodynamic system, whether it undergoing a phase transition also has the similar symmetry change with the phase transition of a normal thermodynamic system?

With the above analyze, we can see that the symmetry will change while the Gauss-Bonnet AdS black hole undergos a phase transition. For the Gauss-Bonnet AdS black hole with $T<T_c$ and $\phi_1$, the black hole moleculars are strongly affected by $\alpha$ and they are generated a certain orientation, which indicates they are in the relative ordered state and have a lower symmetry. While for the Gauss-Bonnet AdS black hole with the another phase $\phi_2$ and the same temperature $T<T_c$, the effect of $\alpha$ on black hole moleculars becomes less powerful. The order of black hole moleculars is relative decreasing, and the black hole has a higher symmetry. With the increase of temperature, the intense thermal motion of black hole moleculars makes the order of black hole moleculars weaken. Especially, when the temperature is more than the critical value $T_c$, the thermal motion of black hole moleculars leads to the order of black hole moleculars be zero. Note that for the Gauss-Bonnet AdS black hole with the lower temperature $T<T_c$, the black hole moleculars have a lower symmetry and higher order, and the order parameter $\phi(T)$ is not equal to zero. While the black hole moleculars have a higher symmetry and lower order, and the order parameter $\phi(T)$ is zero for the Gauss-Bonnet AdS black hole with $T>T_c$.

In the following, we will give the critical exponents. From the Landau's opinion, the order parameter $\phi(T)$ is a small amount near the critical temperature $T_c$. And the Gibbs function $G(T,\phi)$ can be expanded as the power of $\phi(T)$ near the critical temperature $T_c$. Since the phase transition is due to the order change of black hole moleculars, the system is symmetric under the transform $\phi\leftrightarrows-\phi$. Therefore, the expanded express of the Gibbs function $G(T,\phi)$ as the perturbation series of the order parameter $\phi$ only have the even power terms of $\phi$, no the odd power terms of $\phi$:
\begin{eqnarray}
G(T,\phi)=G_0(T)+\frac{1}{2}a(T)\phi^2+\frac{1}{4}b(T)\phi^4+\cdots,\label{G}
\end{eqnarray}
where $G_0(T)$ is the Gibbs function as $\phi(T)=0$. The form of $\phi(T)$ can be confirmed by the condition of Gibbs function minimum value for the stable equilibrium system with unchanged temperature and pressure. Note that $\phi$ in the function $G(T,\phi)$ is a not independent variable. With the requirement of Gibbs function $G(T,\phi)$ minimum value, there are three solutions:
\begin{eqnarray}
\phi=0,~~~~~~\phi=\pm\sqrt{\frac{-a}{b}}.
\end{eqnarray}
The solution $\phi=0$ stands for the unordered state, which is responding to the system with $T>T_c$ and $a>0$. While the non-zero solution represents the ordered state, which is responding to the system with $T<T_c$ and $a<0$. Since the order parameter $\phi$ changes continuously from zero to non-zero, the parameter a should be zero at $T=T_c$.

For the real order parameter $\phi$, we can simply adopt the parameter $a$ near the critical point as
\begin{eqnarray}
a=a_0\left(\frac{T-T_c}{T_c}\right)=a_0t,~~~~~a_0>0.
\end{eqnarray}
Because of the system with $T<T_c$ leading to $a<0$, thus we generally give the limited condition of $b>0$. From the above analyze, we have
\begin{eqnarray}
\phi=\left\{
  \begin{array}{ll}
   0& ~~~~~~~~~~~\text{for}~~~~~ t>0 \\
    \pm\left(\frac{a_0}{b}\right)^{1/2}(-t)^{1/2} & ~~~~~~~~~~~\text{for}~~~~~ t<0
  \end{array}\right.,\label{phi}
\end{eqnarray}
and the critical exponent $\beta$ equals 1/2.

With the above equation (\ref{phi}), the Gibbs function (\ref{G}) can be rewritten as
\begin{eqnarray}
G(T,\phi)=\left\{
  \begin{array}{ll}
    G_0(T) & ~~~~~~~~~~~\text{for}~~~~~ T>T_c \\
    G_0(T)-\frac{a_0^2}{4b}\left(\frac{T-T_c}{T_c}\right)^2 & ~~~~~~~~~~~\text{for}~~~~~ T<T_c
  \end{array}\right..
\end{eqnarray}
From the express of the heat capacity $C=-T\left(\frac{\partial^2G}{\partial T^2}\right)$, we find the heat capacity at the critical point is jumping, and it has the the following form
\begin{eqnarray}
C(T<T_c)\mid_{T=T_c}-C(T>T_c)\mid_{T=T_c}=\frac{a_0^2}{2bT_c}.\label{C}
\end{eqnarray}
Therefore, the jump of the heat capacity at the critical point exhibits the $\lambda$-like shape. That indicates the heat capacity for the ordered phase is bigger than that for the unordered phase, and the change of heat capacity at the critical point is limited. The critical exponent satisfies $\alpha=\alpha'=0$.

With a unchanged pressure, the total differentiation of the Gibbs function $G(T,\phi)$ reads
\begin{eqnarray}
dG=-SdT-\alpha d\Psi.
\end{eqnarray}
From the equation (\ref{phiT}), the differentiation of the order parameter $\phi$ reads
\begin{eqnarray}
d\phi=\frac{d\Psi}{\chi(\Psi_c-3/8\pi)}.
\end{eqnarray}
Considering the above equation and (\ref{G}), we have
\begin{eqnarray}
-\left(\frac{\partial\phi}{\partial\alpha}\right)_T=\frac{\chi(\Psi_c-3/8\pi)}{a+3b\phi^2}
=\left\{
  \begin{array}{ll}
    \frac{\chi(\Psi_c-3/8\pi)}{a_0t} & ~~~~~~~~~~~\text{for}~~~~~ t>0 \\
    \frac{\chi(\Psi_c-3/8\pi)}{-2a_0t}& ~~~~~~~~~~~\text{for}~~~~~ t<0
  \end{array}\right..
\end{eqnarray}
Thus, the critical exponent $\gamma=\gamma'=1$. Since the Gauss-Bonnet coefficient $\alpha$ is proportional to the three powers of the order parameter $\phi$, the critical exponent $\delta=3$, which is consistent with the result in Ref. \cite{Kubiznak2012,Dehyadegari2017,Wei2013}. From the point of view of entropy, the unordered state of the Gauss-Bonnet AdS black hole is of $S=S_0$, while the ordered state is $S=S_0+\frac{a_0^2t}{2bT_c}$. For the case of $t=0$, the entropy of the ordered state is equal to that of the unordered state. That indicates the entropy of black hole is continuous at the critical point.

With the above analyzes, we point out for the Gauss-Bonnet AdS black hole with the temperature ($T<T_c$), the phase transition is the order-unorder one, which is due to the black hole moleculars affected by the Gauss-Bonnet parameter $\alpha$. These results will further expand our understanding of the black hole molecules.

\section{Thermodynamic geometry of the Gauss-Bonnet AdS Black Hole}
\label{scheme5}

In the last part, we have given the parameters a and b in the equation (\ref{G}), which are related with the black hole property. However, the critical exponents are all independent with a and b, as well as a normal thermodynamic system. The reason is that the fluctuation of the order parameter $\phi$ near the critical point is neglected when we analyze the continuous phase transition. In Refs. \cite{Ruppeiner2008,Ruppeiner1995}, the authors investigated the phase transition structure of black holes through the singularity of the spacetime scalar curvature. Thus we can investigate the scalar curvature of the Gauss-Bonnet AdS black hole to reveal the microstructure of the black hole molecules.

\begin{figure*}[htb]
\begin{center}
\includegraphics[width=6.5cm,height=6cm]{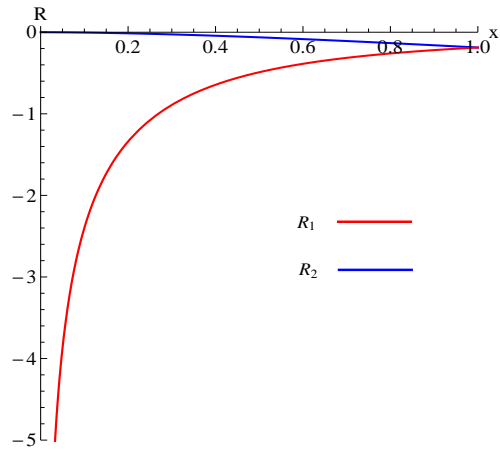}
\end{center}
\vskip -4mm \caption{The Ricci scalar with different radios of black hole horizon radiuses for the parameter adoption $3\sqrt{6}\pi^2\alpha^{3/2}=2$.}\label{R-x}
\end{figure*}

The Ricci scalar based on the Ref. \cite{Miao2018a} reads as
\begin{eqnarray}
R=-\frac{4}{\pi^2 r_+(r_+^2+2\alpha)(8\pi r_+^2P+3)}.
\end{eqnarray}
Since there are two forms of the horizon radius for the black hole with a given temperature $T<T_c$, the Ricci scalar also have two forms (one stands for the order parameter $\phi_1$, another is related with $\phi_2$):
\begin{eqnarray}
R_1&=&-\frac{2(1+4x+x^2)}{3\sqrt{6}\alpha^{3/2}\pi^2\sqrt{x}(1+3x)(1+4x+3x^2)},\label{R1}\\
R_2&=&-\frac{2x^{3/2}(1+4x+x^2)}{3\sqrt{6}\alpha^{3/2}\pi^2(3+x)(3+4x+x^2)}.\label{R2}
\end{eqnarray}
The Ricci scalar plots with different radios of black hole horizon radiuses are given in Fig. \ref{R-x}.

As we known from Refs. \cite{Mirza2008,Mirza2009}, for anyon gas, if the scalar curvature is positive, the average interaction of particles is repulsive, whereas the average interaction is attractive if the Ricci scalar is negative. Especially, there is no interaction of particles for the case of $R=0$. From the Fig. \ref{R-x}, we can obtain the relation $0>R_2>R_1$, which means the average interaction of the black hole molecules for the phase with the order parameter $\phi_2$ is less than the one with the order parameter $\phi_1$.

From the express of the density of black hole molecules $n=\frac{N}{V}=\frac{3}{\gamma l_p^2r_+}$, it is clearly that the density for the phase with the order parameter $\phi_2$ is less than the one with the order parameter $\phi_1$. And from the equations (\ref{R1}) and (\ref{R2}) the values of the Ricci scalar for both two phases are both increasing with the Gauss-Bonnet coefficient $\alpha$ until $R\rightarrow0$. For the fixed temperature and presser, the Gauss-Bonnet coefficient $\alpha$ will increase with the increasing of the black hole horizon radius, while the density and the interaction of black hole molecules will decrease. Therefore, we hold that the Gauss-Bonnet coefficient $\alpha$ plays two roles in a phase transition: one is changing the order of black hole molecules, another is changing the density of black hole molecules. That is just the main reason of phase transition for the Gauss-Bonnet AdS black hole.

\section{discussions and conclusions}
\label{scheme6}
Black hole physics, especially the black hole thermodynamic which is directly involving gravitation, statistics, particle, the field theory and so on, have attracted much attention. Especially, the black hole thermodynamic plays an important role \cite{Bekenstein1972,Bekenstein1974,Bekenstein1973,Bardeen1973,Hawking1974,Hawking1975}. Although the precise statistical description of the corresponding thermodynamic states of black holes is still unclear, the study of the thermodynamic properties and critical phenomenon of black holes is always a concerned issue.

In this paper, we adopted different independent dual parameters to explore the phase transition of the Gauss-Bonnet AdS black hole through the Maxwell's equal-area law. It has been shown that the phase transition point with a given temperature $T<T_c$ is the same for the three adoptions ($P-V,~T-S,~\alpha-\Psi$), while it is different for the adoption of $P-v$. Since the phase transition of black hole with the same condition is independent with the concrete physical process, the parameters $P-v$ are not regarded as the independent dual parameters of the Gauss-Bonnet AdS black hole. This result will be provide the theoretical basis of adopting independent parameters to explore the critical phenomenon of different AdS black holes.

Because of the similarity between the phase transition of the Gauss-Bonnet AdS black hole and that of a vdW system, we have assumed from the microcosmic level a black hole is made up of black hole molecules, which are carrying the message of entropy. The results shown that the phase transition with a certain temperature $T<T_c$ is determined by the radio between $\sqrt{\alpha}$ and the horizon radius, is not only the one from a small black hole to a big one. Therefore, we introduced a new order parameter $\phi(T)$ to investigate the phase transition of the Gauss-Bonnet AdS black hole. Furthermore, the critical exponents have been given in the part \ref{scheme5}.

Finally, we investigated the microstructure of black hole molecules by the spacetime scalar curvature. Since the Schwarzschild AdS black hole is made up of the uncharged black hole molecules, the Ricci scalar is negative, so is the Gauss-Bonnet AdS black hole (see Fig. \ref{R-x}). For the Gauss-Bonnet AdS black hole with the certain temperature and pressure ($T<T_c$, $P<P_c$), the Ricci scalars at the boundary of the two-phase coexistence area are different, that is due to the different values of the order parameter $\phi$ at the boundary of the two-phase coexistence area. The average interaction of black hole molecules of the uncharged Gauss-Bonnet AdS black hole is attractive and it will be close to zero when the Gauss-Bonnet coefficient $\alpha$ is increasing.

This work reflected the microstructure of the Gauss-Bonnet AdS black hole, that will be provide the certain help to explore deeply the microstructure of a black hole, especially understand the basic gravity property of black hole. In particular, the in-depth study of the black hole microscopic structure will help to understand the basic properties of black hole gravity, and it will also have very important value for the establishment of quantum gravity.

\section*{Acknowledgements}
We would like to thank Prof. Zong-Hong Zhu and Meng-Sen Ma for their indispensable discussions and comments. This work was supported by the Natural Science Foundation of China (Grant No. 11705106, Grant No. 11705107 and Grant No. 11475108).

\end{document}